\documentclass[aps,pre,a4paper,10pt,twocolumn,showkeys,reprint,floatfix]{revtex4-1}
\usepackage[utf8]{inputenc}
\usepackage[T1]{fontenc}
\usepackage{amsmath}
\usepackage{amssymb}
\usepackage{bm}
\usepackage{graphicx}
\usepackage{csquotes}
\usepackage[colorlinks=true,hyperfootnotes=true,breaklinks=true]{hyperref}
\usepackage{color,soul}

\setstcolor{red}
\setulcolor{green}

\begin{document}

\title{Perfect cycles in the synchronous Heider dynamics in complete network}

\author{Zdzisław~Burda}
\email{zdzislaw.burda@agh.edu.pl}
\thanks{\includegraphics[scale=0.1]{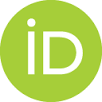}~\href{https://orcid.org/0000-0003-4030-6955}{0000-0003-4030-6955}}
\affiliation{\href{http://www.agh.edu.pl/}{AGH University of Science and Technology, Faculty of Physics and Applied Computer Science, al. Mickiewicza 30, 30-059 Kraków, Poland}}

\author{Małgorzata J. Krawczyk}
\email{gos@agh.edu.pl}
\thanks{\includegraphics[scale=0.1]{ORCID.png}~\href{https://orcid.org/0000-0002-9656-9570}{0000-0002-9656-9570}}
\affiliation{\href{http://www.agh.edu.pl/}{AGH University of Science and Technology, Faculty of Physics and Applied Computer Science, al. Mickiewicza 30, 30-059 Kraków, Poland}}

\author{Krzysztof Kułakowski}
\email{krzysztof.kulakowski@agh.edu.pl}
\thanks{\includegraphics[scale=0.1]{ORCID.png}~\href{https://orcid.org/ 0000-0003-1168-7883}{ 0000-0003-1168-7883}}
\affiliation{\href{http://www.agh.edu.pl/}{AGH University of Science and Technology, Faculty of Physics and Applied Computer Science, al. Mickiewicza 30, 30-059 Kraków, Poland}}


\begin{abstract}
We discuss a cellular automaton simulating the process of reaching Heider balance
in a fully connected network. The dynamics of the automaton is defined by a deterministic,
synchronous and global update rule. The dynamics has a very rich spectrum of attractors
including fixed points and limit cycles, the length and number of which change with 
the size of the system. In this paper we concentrate on a class of limit cycles 
that preserve energy spectrum of the consecutive states. We call such limit cycles perfect.
Consecutive states in a perfect cycle are separated from each other by the same 
Hamming distance. Also the Hamming distance between any two states
separated by $k$ steps in a perfect cycle is the same for all such pairs of states. 
The states of a perfect cycle form a very symmetric trajectory 
in the configuration space. We argue that the symmetry of the trajectories
is rooted in the permutation symmetry of vertices of the network 
and a local symmetry of a certain energy function measuring the level of
balance/frustration of triads.

\end{abstract}

\date{\today}

\maketitle

\section{Introduction}
We study dynamics of spin variables $\pm 1$ defined on
edges of a complete graph on $N$ nodes. 
The spins change in discrete time according to 
the following synchronous update rule \cite{kkb}
\begin{equation}
s_{ij}(t+1)=\textrm{sign} \sum_{k \ne i,j} 
s_{ik}(t) s_{kj}(t) \quad \forall i\ne j ,
\label{eq:s_sss}
\end{equation}
Single indices $i,j,k \in \{$1,\ldots, N$\}$ refer to nodes. 
Pairs of indices, like $ij$, refer to edges. Edges are
undirected so $ij$ is equivalent to $ji$. There are no
self-connections so by default $s_{ii}=0$. For convenience we 
assume that $N$ is odd. This implies
that the sum on the right hand side of (\ref{eq:s_sss})
is strictly positive or negative. It is never zero.

The dynamics (\ref{eq:s_sss}) is motivated by the idea of 
the Heider balance \cite{h} 
in social networks, where the variables $s_{ij}=\pm 1$
represent relationships between agents represented by 
nodes $i$ and $j$ of the graph. The relationships 
can be either friendly $(+1)$ or hostile $(-1)$. They are
assumed to symmetric: $s_{ij}=s_{ji}$. 

This kind of dynamics is known to generically lead to a 
final state where the system divides into two groups 
\cite{akr1,kgg,akr2,mkks,mwk} internally friendly but mutually hostile. 
Such states are termed 'balanced' 
\cite{h}. Here we abstract from the sociological interpretation \cite{h} 
and focus on mathematical properties of the dynamics itself.
We are mainly interested in final states reached during the 
evolution. In addition to 'balanced' states,
which are fixed points of the dynamics, the dynamics can lead 
to jammed states, which are also fixed points but 
they are not balanced \cite{akr1}. 
More interestingly, the dynamics also has limit cycles of 
different lengths. The fixed points and
limit cycles can be used to classify states by basins of attraction they 
belong to. The statistics of basins of attraction for small systems 
was reported in \cite{kkb}. The aim of the present paper is to 
explore properties of limit cycles, in particular of perfect limit cycles 
to be defined below.

\section{Observables}

Let's introduce quantities that are useful 
in probing the behaviour of the system. 
It is convenient to define an energy function
\begin{equation}
U = -\sum _{i<j<k} s_{ij} s_{jk} s_{ki} = \sum_{i<j<k} u_{ijk} 
\label{U}
\end{equation}
where $u_{ijk} = -s_{ij} s_{jk} s_{ki}$ is energy of
triangle $ijk$. The triangle energy is $-1$ when the triad $ijk$ 
is balanced and $+1$ when it is frustrated. Because edges are undirected,
any permutation of indices $ijk$ corresponds to the same triangle. 
A balanced state consists only of balanced 
triads. Energy of a balanced state is $U_{min}=-\binom{N}{3}$. 
This is a global minimum of the energy function. 
A fully frustrated state has the energy equal 
$U_{max}=\binom{N}{3}$. A fully frustrated state can be obtained
from a balanced state by flipping all spins 
$s_{ij} \rightarrow - s_{ij}$. One can also define edge energy
as a sum of energies of all triangles sharing the edge
\begin{equation}
u_{ij}= \sum_{k} u_{ijk}
\label{eq:een}
\end{equation}
and similarly node energy as a sum of energies of all triangles
sharing the node
\begin{equation}
u_i = \sum_{j<k} u_{ijk} .
\label{eq:nen}
\end{equation}
Clearly $\sum_i u_i= \sum_{i<j} u_{ij} = 3 U$.
Each triangle energy configuration $\{u_{ijk}\}_{i<j<k}$ has 
a $2^{N-1}$-fold 
degeneration meaning that there are $2^{N-1}$ distinct
spin configurations having the same triangle energies. 
One can obtain them from each other by flipping all spins 
sharing a node. This operation does not change triangle
energies because it flips an even number of spins in each
triangle. This is a local gauge symmetry of the system. 
This operation can be repeated for $N-1$ nodes, leading to 
$2^{N-1}$ different spin configurations for every triangle 
energy configuration. Note that the initial 
configuration would be restored, if the gauge 
transformation was repeated for all $N$ nodes. Therefore 
'gauge orbits' consist of $2^{N-1}$ and not $2^N$ different
spin configurations. We can define energy spectra:
triangle energy spectrum $n_t(u)$ is the number of
triangles having energy $u$, edge energy spectrum 
$n_e(u)$  is the number of edges having energy $u$, 
and node energy spectrum $n_n(u)$ is the number of nodes
having energy $u$. Formally we can write 
$n_t(u) = \sum_{i<j<k} \delta_{u,u_{ijk}}$,
$n_e(u) = \sum_{i<j} \delta_{u,u_{ij}}$ 
$n_n(u) = \sum_{i} \delta_{u,u_i}$
where $\delta_{a,b}$ is the Kronecker delta. The energy 
spectra take nonzero values from the range 
$\pm 1$ for triangles, $\pm (N-2)$ for edges and
$\pm (N-1)(N-2)/2$ for nodes. 

The proximity of spin configurations $A$ and $B$ 
can be measured by the Hamming distance
\begin{equation} \label{eq:dH}
    d_H(A,B) = \frac{1}{4} \sum_{i<j} 
    \left(s_{ij}(A) - s_{ij}(B)\right)^2 \ .
\end{equation}
Similarly one can define the Hamming distance between
triad configurations $\{u_{ijk}(A)\}$ and $\{u_{ijk}(B)\}$
\begin{equation}
    D_H(A,B) = \frac{1}{4} \sum_{i<j<k} 
    \left(u_{ijk}(A) - u_{ijk}(B)\right)^2 .
\end{equation}
since triangle energies $u_{ijk}$'s are also binary variables.
The Hamming distance $D_H(A,B)$ is equal zero 
for $A$ and $B$ from the set of $2^{N-1}$ spin configurations 
having the same triangle energies. It does not imply that
$A=B$ so $D_H$ is not a distance 
for spin configurations. Obviously $d_H(A,B)=0$ 
implies that $D_H(A,B)=0$, but not vice versa. We shall write
$A \asymp B$ if $D_H(A,B)=0$, to denote gauge equivalent configurations.

We can use the Hamming distance (\ref{eq:dH}) to measure proximity
of consecutive configurations 
$A_0 \rightarrow A_1 \rightarrow A_2 \rightarrow \ldots $ 
generated by the synchronous dynamics (\ref{eq:s_sss}) and in particular 
to find fixed points and limit cycles of the dynamics. A configuration $A_t$
such that  $d_H(A_t,A_{t+1})=0$ is a fixed point 
of the dynamics. The minimal value $c$ such that $d_H(A_t,A_{t+c})=0$ 
is the length of a limit cycle. The corresponding cycle consists of configurations
$A_t, A_{t+1},\ldots, A_{t+c-1}$.  Initial configurations $A_0$ of any sequence
of configurations $A_0\rightarrow A_1 \rightarrow \ldots$ generated by the dynamics 
(\ref{eq:s_sss}) can be classified by 
a fixed point or limit cycle of the sequence. With a limit cycle (or a fixed point) 
one can associate a basin of attraction that is a set of initial states $A_0$ 
which lead to this limit cycle.

The update rule (\ref{eq:s_sss}) can be written in the following way
\begin{equation}
s_{ij}(t+1)= - \textrm{sign}\left( s_{ij}(t) u_{ij}(t) \right) = \left\{ 
\begin{array}{rl}-s_{ij}(t)  &   \mbox{if} \ u_{ij}(t)>0 \\  
                            s_{ij}(t) &  \mbox{if} \ u_{ij}(t)<0 
\end{array} \right. 
\end{equation}
If this update rule was applied asynchronously that is to one edge at one time,
it would never increase energy, and it would drive the system to a local energy minimum. 
We are however interested in synchronous dynamics. In this case more than one edge 
of a triangle can be updated simultaneously and in effect triangle energy and thus also energy of 
the system can increase. The number of spins flipped in one step of synchronous dynamics (\ref{eq:s_sss}) is equal to the number of positive $u_{ij}$'s, so 
\begin{equation}
\label{eq:dn}
    d_H(A_t,A_{t+1}) = \sum_{i<j} \Theta\left(u_{ij}(t)\right) = \sum_{u>0} n_e(u,t)
\end{equation}
where $\Theta$ is the Heaviside step function, and $n_e(u,t)$ is the edge energy spectrum
of the configuration $A_t$. It follows that $A_t$ is a fixed point of the dynamics,
if all edge energies are negative, that is $n_e(u,t)=0$ for $u>0$.  The edge spectrum is said to be steady
for $t>t_0$ if  $n_e(u,t)=n_e(u,t+1)$ for all $u$ and $t>t_0$. This just means that the spectrum does not change for $t>t_0$. For steady spectra the time dependence can be skipped $n_e(u,t)=n_e(u)$. Fixed points have steady spectra, but as we will see also some cycles have. We will call such cycles {\em perfect}. The Hamming distance between any two consecutive configurations of a perfect cycle is constant: $d_H(A_t,A_{t+1}) = {\rm const}$, as follows from (\ref{eq:dn}). In the next section we will discuss examples of perfect cycles.

\section{Perfect cycles}

\begin{figure}
\begin{center}
\includegraphics[width=\columnwidth]{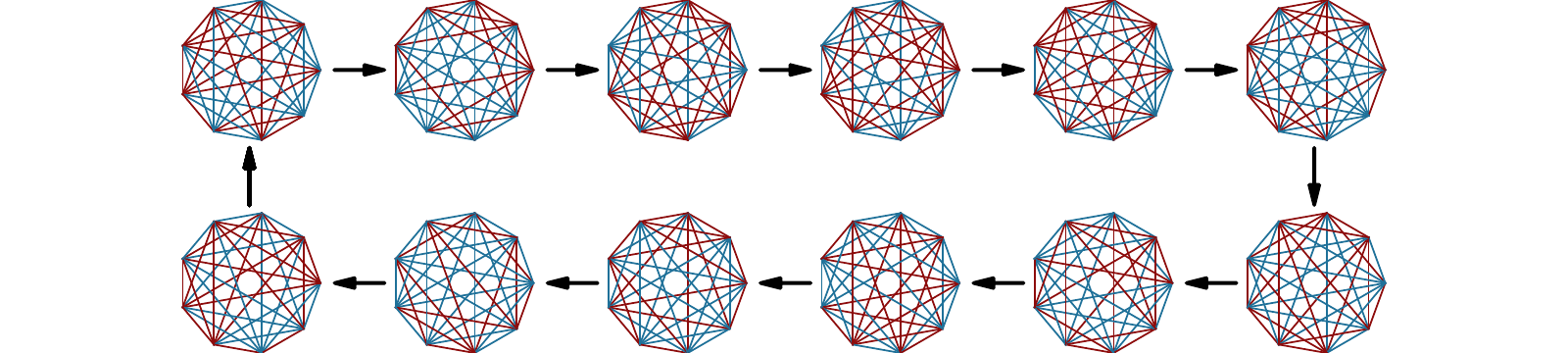}
\caption{An example of a perfect limit cycle on a complete graph on $N=9$ nodes. The cycle consists of $c=12$
configurations. They are drawn in the order they appear in the cycle. 
Edges for $s_{ij}=1$ are in red and for $s_{ij}=-1$ in blue. Labels of vertices are not displayed. 
Vertices are numbered $i=1,\ldots, 9$ clockwise, starting from the vertex $1$ on the top. The configuration 
in the left upper corner is equivalent to that given by the matrix (\ref{eq:A0}).}
\label{c12}
\end{center}
\end{figure}

Let us first consider the system for $N=9$. This is a good
test site because the update rule (\ref{eq:s_sss})
can be applied to all $2^{36}$ spin configuration using a computer program, so one can test
all configurations.
Already for $N=11$ the number of configurations is too large for 
an exhaustive computation for all configurations. We found that there are  
$967680$ cycles of length $c=12$ for $N=9$. An example of a configuration
belonging to a perfect cycle is
\begin{equation}
s = \left(
{\footnotesize
\begin{array}{rrr rrr rrr} 
0   & -1 & -1 & -1 & +1 & -1 & +1 & +1 & -1 \\
 -1 & 0  & -1 & -1 & +1 & -1 & +1 & +1 & +1 \\
 -1 & -1 & 0  & -1 & -1 & +1 & +1 & +1 & -1 \\
 -1 & -1 & -1 &  0 & +1 & +1 & -1 & -1  & -1 \\
 +1 & +1 & -1 & +1 & 0 & -1 &  -1 & -1 & +1\\
 -1 & -1 & +1 & +1 & -1 & 0 & +1 & -1 & -1 \\
 +1 & +1 & +1 & -1 & -1 & -1 & +1 & 0 & +1\\
 -1 & +1 & -1 & -1 & +1 & -1 & +1 & +1 & 0  
\end{array}}
\right) \label{eq:A0}
\end{equation}
A graphical representation of this state and remaining states belonging
to the perfect cycle is shown in Fig. \ref{c12}. With a naked 
eye it is rather difficult to see what makes 
these states form a perfect cycle. The situation changes when the energy spectra of these states are analysed,
because then you can observe that all states have constant spectra.
Edge energy spectrum is given in Table \ref{t:N9c12}.
One can easily see that energy of the states is
$U= \frac{1}{3} \sum_u u n_e(u) = -6$,
and the distance between any two consecutive states in
the cycle (\ref{eq:dn}) is $d_H(A_t,A_{t+1}) = \sum_{u>0} n_e(u) = 18$.
\begin{table}
 \begin{center}
\begin{tabular}{ |c|c|c|c|c|c|c|c|c|}
 \hline
 $u$  & -7 & -5 & -3 & -1 & +1 & +3 & +5 & +7  \\
 \hline
 $n_e(u)$ &  3 & 2 & 4 & 9 & 12 & 4 & 2 & 0  \\ 
 \hline
\end{tabular}
\caption{ Edge spectrum of states belonging to the perfect cycle
of length $c=12$ for $N=9$. \label{t:N9c12}}
\end{center}
\end{table} 
Using a computer program we have checked that 
configurations separated by two steps in the cycle differ by a constant number of spins
$d_H(A_t,A_{t+2})=22$. Similarly, the distance between any two
configurations separated by three steps is constant 
$d_H(A_t,A_{t+3})=20$. Generally we found that for any
$s$ the distance $d_H(A_t,A_{t+s})$ in the cycle is constant  for all $t$ 
as long as $s$ is fixed. For completeness, $d_H(A_t,A_{t+s})=10,18,20$, for $s=4,5,6$. 
Also, $d_H(A_t,A_{t+s})$ is the same as for $s \rightarrow 12 \pm s$. The plus minus
symmetry follows from the symmetry of the distance function:
$d_H(A_t,A_{t+s})=d_H(A_{t-s},A_t)=d_H(A_t,A_{t-s})$.

Also the number of triangles by which $A_t$ and $A_{t+s}$ differ 
is constant for all $t$ when $s$ is fixed, and it is $D_H(A_t,A_{t+s}) = 36,32,16,32,36,0$ 
for $s=1,2,3,4,5,6$.

Let us also mention some other features that are present for all such cycles.
There are six types of edges which differ in the sequence of states:
(1) three links remain constant ($+1$ or $-1$) with energy 
$E_{ij}=-s_{ij}\sum_ks_{ik}s_{kj}$ equal to -7; (2) four links change $12$ times, with $E_{ij}=+1$; 
(3) six links change $4$ times ($s_{ij}=1,1,1,-1,-1,-1$ and cyclically, 
$E_{ij}=-1,-5,+1$ and cyclically); (4) three links change $8$ times ($s_{ij}=-1,-1,+1$ and cyclically, $E_{ij}=-1,+3,+3$ and cyclically); 
(5) eight links change $6$ times ($s_{ij}=1,1,-1,-1$ and cyclically, 
$E_{ij}=-1,+1$ and cyclically); (6) twelve links change $6$ times ($s_{ij}=-1,-1,1,1,-1,1,1,1,-1,-1,1,-1$, $E_{ij}=-1,+5,-3,+1,+3,-3$ and cyclically.

Positions of states in the cycle are in equivalent in the sense that measuring relative
changes to other states in the cycle you are not able to distinguish the states.
This equivalence must be rooted in a symmetry of the system.  
There are two basic symmetries that should be taken into account: 
the automorphism of the complete graph, that is equivalent to 
the permutation of indices of the complete graph, and the gauge symmetry 
of spin configurations which preserves triangle energies. The hypothesis is that
every configuration of a perfect cycle can be obtained from the previous one by
a permutation of indices and a gauge transformation of spins. This in turn means
that there exists a permutation of indices $\pi$ of
indices such that $A_{t+1} \asymp \pi(A_t)$. We have tested this hypothesis
for $N=9$. Applying the update rule to the state
(\ref{eq:A0}), that we denote by $A_t$, we obtained a state $A_{t+1}$. 
Then we have determined all permutations $\pi$'s
such that $A_{t+1} \asymp \pi(A_t)$ by checking if the condition 
\begin{equation}
D_H(A_{t+1},\pi(A_t))=0
\label{eq:ge}
\end{equation}
is fulfilled.We have found that there are eight such permutations:
\begin{equation}
\begin{split}
\pi_1 & = (7,3,2,4,9,1,6,8,5) \\
\pi_2 & = (7,3,2,8,9,1,6,4,5) \\
\pi_3 & = (7,3,9,4,2,1,6,8,5) \\
\pi_4 & = (7,3,9,8,2,1,6,4,5) \\
\pi_5 & = (7,5,2,4,9,1,6,8,3) \\
\pi_6 & = (7,5,2,8,9,1,6,4,3) \\
\pi_7 & = (7,5,9,4,2,1,6,8,3) \\
\pi_8 & = (7,5,9,8,2,1,6,4,3) 
\end{split} . \label{eq:pi8}
\end{equation}
Renaming $A_{t+1}$ to $A_t$ and applying equation (\ref{eq:ge}) we have  
again found the same eight permutations. It turns out, that the same eight
permutations map any configuration onto the next configurations in the cycle. 
The permutations can be determined by exhaustive search but such a procedure 
is inefficient because there are $9!$ permutations. 
One can improve the search using the information encoded in the node energy lists of the 
configurations of the cycle, see Table \ref{t:ne}. By analysing migration of node energies in consecutive configurations of the cycle one can learn about the corresponding permutations of indices which fulfill the condition (\ref{eq:ge}). For example, energy $10$ moves from 
the position  $1$ to $7$, from $7$ to $6$ and from $6$ to $1$. This means that 
the permutation has a cycle $(7,6,1)$. This in turn reduces the number of remaining 
permutations to $6!$. Further, by analysing migration of remaining items row by row in Table \ref{t:ne}, 
one can find other cycles and reconstruct  all the permutations (\ref{eq:pi8}). For completeness we give
the cycle decomposition of the permutations:
$\pi_1 = (7,6,1)(3,2) (9,5)$, $\pi_2=(7,6,1)(3,2)(8,4)(9,5)$,
$\pi_3=(7,6,1)(3,9,5,2)$, $\pi_4=(7,6,1)(3,9,5,2)(8,4)$, $\pi_5=(7,6,1)(5,9,3,2)$,
$\pi_6=(7,6,1)(5,9,3,2)(8,4)$, $\pi_7=(7,6,1)(5,2)(9,3)$ and
$\pi_8=(7,6,1)(5,2)(9,3)(8,4)$.
We can use the result to calculate the number of the corresponding cycles. Each cycle is represented by
twelve tables like Table \ref{t:ne}. Twelve tables which differ by a cyclic permutation of rows are equivalent, since
for a cycle it does not matter which configuration is listed first.
Any permutation of columns (nodes) produces a  table with the same node energy spectrum but possible
with different positions on the lists. The tables obtained by $9!$ permutations generically correspond to different 
cycles, but not always. One has to take into account that eight permutations (\ref{eq:pi8}) 
produce a cyclic shift of rows in the table, as follows from the fact that
the effect of the these permutations is equivalent to applying one step the dynamics
(\ref{eq:s_sss}). Thus permutations of indices generate  $9!/8/12$ non-equivalent tables.
Due to the gauge symmetry, each energy configuration is realised by
$2^8$ distinct spin configurations. Putting theses factors together, we find that there are 
\begin{equation}
\frac{9!}{8\cdot 12} 2^8 = 967680
\label{eq:enum}
\end{equation} 
distinct perfect cycles having the node energy spectrum $n_n(10)=1$, $n_n(-2)=5$, $n_n(-6)=3$. 
We have confirmed this prediction numerically by checking the effect of the action of the transformation 
(\ref{eq:s_sss}) for all configurations for $N=9$. We also found that 
there are no other cycles of length $c=12$ for $N=9$.
\begin{table}
 \begin{center}
\begin{tabular}{ |r|rrr rrr rrr |}
 \hline
   & 1 & 2 & 3 & 4 & 5 & 6 & 7 & 8 & 9 \\ \hline
1 &   10&-2&-2&-6&-2&-6&-2&-6&-2 \\ \hline
2 &   -6&-2&-2&-6&-2&-2&10&-6&-2 \\ \hline
3 &   -2&-2&-2&-6&-2&10&-6&-6&-2 \\ \hline
4 &  10&-2&-2&-6&-2&-6&-2&-6&-2 \\ \hline
5 &  -6&-2&-2&-6&-2&-2&10&-6&-2 \\ \hline
6 &  -2&-2&-2&-6&-2&10&-6&-6&-2 \\ \hline
7 &  10&-2&-2&-6&-2&-6&-2&-6&-2 \\ \hline
8 &  -6&-2&-2&-6&-2&-2&10&-6&-2\\ \hline 
9 &  -2&-2&-2&-6&-2&10&-6&-6&-2 \\ \hline
10 & 10&-2&-2&-6&-2&-6&-2&-6&-2 \\ \hline
11 &  -6&-2&-2&-6&-2&-2&10&-6&-2 \\ \hline
12 &  -2&-2&-2&-6&-2&10&-6&-6&-2 \\ \hline
\end{tabular}
\caption{Each row contains a list of node energies of a configuration of the cycle. 
The row in the table header numbers the nodes, and the column
on the left side numbers the successive cycle configurations.
The node energy spectrum is steady: $n_n(10)=1$, $n_n(-2)=5$ and $n_n(-6)=3$.  \label{t:ne}}
\end{center}
\end{table}
The perfect cycles of length $c=12$ for $N=9$ have relatively small basins of attraction which consist of $312$ states
including the $12$ states belonging to the cycle and $300$ other states. $12$ out of $300$ states are mirror
states of those belonging to the cycle. Mirror state $s^*$ of a state $s$ is a state with all opposite signs
$s^*_{ij} = -s_{ij}$ for all $ij$. The remaining $288$ states can be divided into $12$ groups, each having
$12$ pairs of mutually mirror states. Each of the $12$ groups is associated with one state of the cycle to which all $24$ states from the group are transformed in a single step of the dynamics (\ref{eq:s_sss}).  None of $300$  states has a predecessor. Such states are sometimes called 'Garden of Eden' \cite{m}.
 
 We have also studied systems for $N>9$ to search for perfect cycles. In this case,
 however, we performed a random search since as mentioned the
 number of configurations is too large for these systems to be exhaustively browsed.
 We have found perfect cycles of length $c=14$ for $N=11$. The edge energy spectra 
 of these cycles is shown in Table \ref{t:en11c14}. 
 \begin{table}
 \begin{center}
\begin{tabular}{ |c|c|c|c|c|c|c|c|c|c|c|}
 \hline
 $u$ & -9 & -7 & -5 & -3 & -1 & +1 & +3 & +5 & +7 & +9  \\
 \hline
 $n_e(u)$ & 0 & 0 & 4 & 10 & 15 & 13 & 9 & 3 & 1 & 0 \\ 
 \hline
\end{tabular}
\caption{Edge energy spectrum of states belonging to the perfect cycle
of length $c=14$ for $N=11$. \label{t:en11c14}}
\end{center}
\end{table}
 As follows from the table, energy of the configurations
is $U= \frac{1}{3} \sum_u u n_e(u) = -1$,
and the Hamming distance between any two neighbouring states in
the cycle (\ref{eq:dn}) is $d_H(A_t,A_{t+1}) = \sum_{u>0} n_e(u) = 26$.
The corresponding node energy spectrum is $n_n(-9) = 1$, $n_n(-5)=4$,
$n_n(3)=4$,  $n_n(7)=2$ and $n_n(u)=0$ for other values of $u$.
As before we found that $d_H(A_t,A_{t+s})$ and $D_H(A_t,A_{t+s})$
for fixed $s$ are independent of $t$, so all configurations of the cycle are equivalent, and symmetrically distributed in the configuration space. We found that there are two distinct permutations fulfilling the condition (\ref{eq:ge}). They can be decomposed into a cycle of length seven 
and two cycles of length two. Using the same enumeration argument 
as before (\ref{eq:enum}) this gives $11!/2/14\times 2^{10}$ such cycles. One would
need to check all configurations,
to exclude that there are no other cycles (with a different energy spectrum)
for $N=11$.
We have also found a perfect cycle of length $c=12$ for $N=13$.
The edge energy spectrum is given in Table \ref{t:en13c12}.
The energy of the configurations is $U= \frac{1}{3} \sum_u u n_e(u) = -56$,
and the Hamming distance between any two neighbouring states in
the cycle (\ref{eq:dn}) is $d_H(A_t,A_{t+1}) = \sum_{u>0} n_e(u) = 20$.
 \begin{table}
 \begin{center}
\begin{tabular}{ |c|c|c|c|c|c|c|c|c|c|c|c|c|}
 \hline
 $u$ & -11 & -9 & -7 & -5 & -3 & -1 & +1 & +3 & +5 & +7 & +9 & +11  \\
 \hline
 $n_e(u)$ & 1 & 3 & 7 & 8 & 21 & 18 & 12 & 6 & 2 & 0 & 0 & 0\\ 
 \hline
\end{tabular}
\caption{Edge energy spectrum of states belonging to the perfect cycle
of length $c=12$ for $N=13$. \label{t:en13c12}}
\end{center}
\end{table}
The node energy spectrum is $n_n(-20)=4$, $n_n(-16)=2$, $n_n(-12)=4$,
$n_n(-4)=-2$, $n_n(0)=1$. Again we found that $d_H(t,t+s)$ and $D_H(t,t+s)$ are 
 independent on $t$ when $s$ is constant.

\section{Semi-perfect cycles}
Not all limit cycles have steady energy spectra. There are cycles
whose spectra change periodically. We will call them semi-perfect.
As an example let us discuss a semi-perfect cycle that we have 
found for $N=13$. The cycle is representative for all semi-perfect 
cycles in that that it has typical features, but additionally it is the 
longest limit cycle we have found so far.  
It has the length of  $c=48$.  The
energy spectra of the states in the cycle change with the period three.
The edge spectra of three consecutive states of the cycle 
are given in Table \ref{t:en13c48}. 
\begin{table}
 \begin{center}
\begin{tabular}{ |c|c|c|c|c|c|c|c|c|c|c|c|c|}
 \hline
 $u$ & -11 & -9 & -7 & -5 & -3 & -1 & +1 & +3 & +5 & +7 & +9 & +11  \\
 \hline
 $n_{e}(u)$ & 2 & 0 & 3 & 11 & 17 & 15 & 17 & 8 & 4 & 1 & 0 & 0\\ 
 \hline
 $n_{e}(u)$ & 2 & 2 & 6 & 8 & 10 & 16 & 16 & 17 & 1 & 0 & 0 & 0\\ 
 \hline
 $n_{e}(u)$ & 2 & 0 & 2 & 9 & 18 & 15 & 20 & 9 & 2 & 0 & 1 & 0\\ 
 \hline
 \end{tabular}
\caption{Edge energy spectra of three consecutive states belonging 
to the semi-perfect cycle of length $c=48$ for $N=13$. \label{t:en13c48}}
\end{center}
\end{table}
Energies of the states are $U_t=-32,-32,-28$. 
The Hamming distance between neighbouring states is
$d_H(t,t+1) = 34,32,30$ and $D_H(t,t+1)=128,136,134$. 
The values repeat every three steps. If we denote the map corresponding
to a single state (\ref{eq:s_sss}) by $s(t+1)= \Phi(s(t))$, then taking
every third configuration is equivalent to $s(t+3) =\Phi(\Phi(\Phi(s(t)))) = \Psi(s(t))$
where the map is a triple composition of $\Phi$: $\Psi=\Phi\circ\Phi\circ\Phi$.
Viewed from this perspective, the semi-perfect cycle of the dynamics defined by the
map $\Phi$ (\ref{eq:s_sss}) is a perfect cycle for $\Psi$. More generally, the class of semi-perfect
cycles is a class of limit cycles which are perfect for a multiple composition
$\Phi \circ \ldots \circ \Phi$ of the original update rule.

\section{Discussion}

The motivation behind the evolution rule (\ref{eq:s_sss}) is that it locally maximises the number of balanced
triads. Indeed, when performed asynchronously, that is one edge at time, the rule never reduces the number
of balanced triads and thus it leads to a state at local maximum, as far as the number of balanced triads is concerned
(equivalent to local energy  minimum (\ref{U})). 
The synchronous version of the evolution (\ref{eq:s_sss}) where all edges
are updated simultaneously has a far more interesting spectrum of attractors: in addition to fixed points it
has limit cycles of different length and of different symmetry. 
Some limit cycles are surprisingly long. For example we found 
a limit cycle of length $c=48$ for $N=13$. In this paper we mostly focused on a class of
limit cycles which preserve the energy spectrum and are represented by symmetric trajectories in the 
configuration space, such that any two states separated by the same number of steps in the perfect cycle 
are separated by the same Hamming distance in the configuration space. We have argued that the symmetry 
of these trajectories is rooted in the automorphism group of the complete graph on which the system is defined and 
in the local gauge symmetry of the energy function (\ref{U}). 
 
There are many open questions. Is it possible to formulate general conditions that would make it possible to judge 
whether a state belongs to a limit cycle, before checking it explicitly by iterating the equation (\ref{eq:s_sss})? 
What is the longest limit cycle and the longest perfect cycle for the complete network for given $N$? 
What is the abundance of such cycles? We know \cite{kkb} that the fraction of initial states which lead to perfect 
limit cycles of length $c=14$ for $N=11$ is about $10^{-6}$, which is much less than the fraction of perfect cycles
$c=12$ for $N=9$ which is $0.004$. We expect that the percentage of states of perfect cycles decreases with
the system size, but it would be good to find an argument about asymptotic behaviour.

Generally, the dynamics we discussed in this paper is of the type $s(t+1) = \Phi(s(t))$. The map $\Phi$ 
given by Eq. (\ref{eq:s_sss}) is just a particular case. One can change the evolution rule. For example
adding a minus sign to the expression on the right hand side of Eq. (\ref{eq:s_sss}) we would obtain a 
system having a tendency to maximize the number of frustrated triads. Of course this evolution would be 
in one-to-one correspondence to the one discussed here as can be seen by replacing states $s$ in one original 
dynamics by mirror states $s^*$ in the new one. But the question about how the attractors of the evolution
depend on the given map $\Phi$ is quite interesting. For example what is the class of maps $\Phi$ which would 
lead to perfect limit cycles? It would be interesting to study symmetry classes for general maps \cite{k2}.

There is some correspondence of the dynamics of the model discussed in this paper and the quenched Kauffman NK model \cite{k1,ack} of time evolution of networks. As we argued in \cite{kkb}, here the number K of incoming links which determine the current state of a node (here: of a link) evolves with the number of degrees of freedom (here: $N^2$) as a square root of this number (here $N$). An important difference is that in our case, there is only one function (given by Eq. (1)) which determines the state of each link in a subsequent time, and not a random (fixed in the quenched model) set of these functions, different for each node. What is similar is the large number of steady states with minimal energy, which in our case is just the number of balanced states, varying with $N$ as $2^{N-1}$. We add that the process of reaching the Heider balance, modeled by Eq. (1), has been termed as 'social mitosis' \cite{wt}.  
Limit cycles in the Kauffman model \cite{bp,bs} are no less
important than fixed points and have 
biological interpretation. Our results indicate that limit cycles can also occur when evolution is deterministic
and identical for all components of the system.


\end{document}